\documentstyle[12pt]{article}

\newcommand{\spone}{0.9}

\newcommand{\singlespace}{\edef\baselinestretch{\spone}\Large\normalsize}

\begin{document}
\hspace*{\fill} PURD-TH-94-10\\
\vspace{0.5in}

\begin{center}
{\bf STABILITY OF FINE TUNED HIERARCHIES IN STRONGLY COUPLED CHIRAL MODELS}\\
{}~\\
{}~\\
{\bf T.E. Clark and S.T. Love}\\
{\it Department of Physics}\\
{\it Purdue University}\\
{\it West Lafayette, IN 47907-1396
}
\end{center}
{}~\\
{}~\\
\begin{center}
{\bf ABSTRACT}\\
\end{center}

{\singlespace

A fine tuned hierarchy between a strongly coupled high energy compositeness
scale and a much lower chiral symmetry breaking scale is a requisite
ingredient in many models
of dynamical electroweak symmetry breaking. Using a nonperturbative
continuous Wilson
renormalization group equation approach, we explore the stability of such a
hierarchy against quantum fluctuations. }
{}~\\
{}~\\
{}~\\


Many of the currently studied models of dynamical
electroweak symmetry breaking include some strongly interacting sector
acting at a high energy scale, $\Lambda > 10 TeV$, which produces an
essentially composite scalar bosonic degree of freedom. In addition,
this dynamics is also supposed to play a non-trivial
role in the electroweak symmetry breaking whose characteristic scale
is much lower; $\Lambda_F \simeq 250 GeV$.
Thus these models require that a significant hierarchy can be
established between these scales.
Included in this class are
strong extended technicolor models$^{[1]}$, models of broken
technicolor$^{[2]}$ and models involving
heavy quark condensation$^{[3-6]}$.

In general, the
hierarchy is achieved by a fine tuning of parameters close to
the critical value for the chiral symmetry breaking. A prototype of this
behavior is exhibited by the Nambu Jona-Lasinio (NJL) model$^{[7]}$,
where a
fine tuning of the four-fermion coupling allows the emergence of a chiral
symmetry breaking scale far below the compositeness scale. A feature of
this particular model which must be generic in any model which exhibits such
a large hiearchy of scales is that the chiral symmetry phase
transition be of second order. That is,
in order for the hierarchy to be maintained and not have the electroweak scale
driven to be of order $\Lambda$, it is necessary that the order parameter
characterizing the chiral transition must remain zero as the theory
is scaled from $\Lambda$ into the infrared until one reaches the electroweak
scale.

If, on the other hand, quantum fluctuations turn the
transition first order at a scale
$e^{-t_0}\Lambda >> \Lambda_F$, then the order parameter will jump
discontinuously to
be of this value and it will be impossible to maintain the hierarchy
all the way down to the electroweak scale. Instead the hierarchy will
destabilize after $t_0$ e-foldings. Such a situation is an
example of the phenomenon studied by Coleman and Weinberg$^{[8]}$.
It is important to recognize that this
question is distinct from that of the naturalness of the fine tuning of
additive quadratic divergences. Even allowing for such fine tunings, this
destabilization of hierarchies could prove problematic for many of the models
of dynamical electroweak symmetry breaking. Thus, as
emphasized by Chivukula, Golden and Simmons$^{[9]}$, it becomes essential
to explore when the hierarchy can be self
consistently maintained and not destroyed by quantum fluctuations.

The minimal NJL model which is used in the minimal
top condensate model$^{[5]}$ is known to exhibit a
second order chiral transition.
However, since
a top quark mass of 174 GeV, as reported by the CDF collaboration$^{[10]}$,
is outside the range of the
minimal model, it appears that if such models are to be
phenomenologically viable, they must necessarily require extensions beyond the
minimal version. Similarly, while the O(4) linear sigma model containing a
single scalar quartic self coupling has been shown to exhibit a mean field
second order chiral transition, the effective dynamics of
strong extended technicolor models could
require an effective Lagrangian containing more than one
scalar self coupling. Thus we
are led to investigate models containing multiple scalar
quartic self couplings.

Following previous work $^{[9,11,12]}$,
we focus on a model possessing a global
chiral $U(2)_L \times U(2)_R$ symmetry which has two
independent scalar quartic self couplings. The model degrees of freedom include
left and right handed chiral fermions $\psi_{iL}$ and $\psi_{iR}$, $i=1,2$,
transforming as the fundamental $(2,0)$ and $(0,2)$ representations
of the $U(2)_L \times U(2)_R$ groups respectively which
further carry the fundamental, $N_C$, representation of an asymptotically
free gauged
symmetry. Throughout our discussion, we shall neglect these gauge interactions
which are considered to be very feeble at these high energy scales. We assume
that, as a consequence of some unspecified dynamics acting at scale $\Lambda$,
the $U(2)_L \times U(2)_R$ chiral symmetry is spontaneously broken.
This symmetry breaking is further assumed to produce a gauge singlet scalar
composite $\Sigma_{ij}$ which has the $U(2)_L \times U(2)_R$ quantum numbers
of the fermion bilinear $\bar{\psi}_{jR} \psi_{iL}$. That is $\Sigma_{ij}$
transforms as the $(\bar{2},2)$ under the chiral group. Its vacuum
expectation
value, $\frac{v}{\sqrt{2}}\delta_{ij}$, can be interpreted as
an order parameter for
the chiral symmetry breaking. Since we are assuming the
chiral symmetry phase transition is second
order, we are led to study a Ginzburg-Landau effective model whose
Euclidean action at scale $\Lambda$ is given by
$$  \everymath={\displaystyle}
   \begin{array}{rcl}
   S[\Sigma, \Sigma^{\dagger},\psi,\bar{\psi};0] &=& \int d^4 x
\{tr (\partial_\mu \Sigma^{\dagger} \partial^\mu \Sigma)
     + \bar{\psi} \gamma\cdot D\psi + V(x,y,0)\\
       && + \frac{\pi}{\sqrt{2}}g(0)(\bar{\psi}_L \Sigma \psi_R +
          \bar{\psi}_R \Sigma^{\dagger}\psi_L) \}  \, ,
     \end{array}
    \eqno{(1)}
$$
where $D_\mu \psi$ is the fermion gauge covariant derivative and
the invariant potential function takes the form
$$
V(x,y,0) = \frac{1}{2}m^2(0) x + \frac{\pi^2}{12}\lambda_1(0)x^2
+ \frac{\pi^2}{6}\lambda_2(0) y \, .
\eqno{(2)}
$$
Here
$x = tr (\Sigma^{\dagger}\Sigma),y = tr(\Sigma^{\dagger}\Sigma)^2$
are the two independent
$U(2)_L \times U(2)_R$ invariants. A Coleman-Weinberg instability is
signalled by the appearance of a non-trivial global minimum of the effective
potential, $V_{eff}(v)$, with
vanishing renormalized mass. If this occurs at the scale $v=e^{-t_0}\Lambda$,
then the phase transition is driven
first order by quantum fluctuations at this scale
and one can technically achieve a
hierarchy of only $t_0$ e-foldings.

The conditions for the appearance of such a non-trivial global
minimum of the effective potential with vanishing renormalized mass
can be elegantly expressed
in terms of the various renormalization group functions by
employing a formalism due to
Yamagishi$^{[13]}$.
In general, the (Yamagishi) stability function is defined as
$$
Y(t_0)=-\frac{12}{\pi^2}e^{4t_0}\frac{dV_{eff}}{dt_0} \, ,
\eqno{(3)}
$$
with $v=e^{-t_0}$. The effective potential is minimized provided
$$
Y(t_0)=0~~;~~\frac{dY}{dt}|_{t=t_0}<0 \,
\eqno{(4)}
$$
and this minimum is a global one provided $V_{eff}(v)<V_{eff}(0)=0$.
Approximating the full effective potential by
the 1-loop perturbation theory improved
version which includes only the leading logarithmic radiative
corrections of the
dimension four operators appearing in the potential function of Eq.~(2),
while further holding the Yukawa coupling fixed,
these Yamagishi conditions for the appearance
of a global minimum at $t_0$ reduce to
$$ \everymath={\displaystyle}
   \begin{array}{rcl}
&&Y(t_0) = [4 \lambda_1 + 4 \lambda_2 + \beta_1(\lambda)
+\beta_2(\lambda)]|_{t=t_0} = 0\\
&&[4 + \beta_1 \frac{\partial}{\partial \lambda_1}
+ \beta_2 \frac{\partial}{\lambda_2}]
(\beta_1 + \beta_2)|_{t=t_0} > 0~~;~~
\lambda_2(t_0) > 0 \\
&&\lambda_1(t_0) + \lambda_2(t_0) < 0 \, ,
\end{array}
\eqno{(5)}
$$
The first two lines of Eq.~(5) are the fixed Yukawa coupling,
1-loop approximation version of Eq.~(4) and
guarentee a minimum at $t_0$, while the
last inequality insures the minimum is a global one.
Note that for other approximation schemes which are still restricted to
include only the marginal, dimension
four operators, the Yamagishi coditions of Eq.~(5) are modified by the
replacements $\beta_i(t)\rightarrow \bar{\beta}_i(t)=\frac{\beta_i(t)}{1+
\gamma(t)}$ and $\lambda_i(t)\rightarrow \bar{\lambda}_i(t)$,
where $\gamma(t)$ is the anomalous dimension of
$\Sigma$ and $\bar{\beta}_i=
\frac{\partial\bar{\lambda}_i}{\partial t}$.
Using the 1-loop perturbative $\beta$ functions, the Yamagishi
stability condition simply reduces to
a curve in the $\lambda_2(t)/g^2(0)-
\lambda_1(t)/g^2(0)$ plane. The signal for the Coleman-Weinberg instability
is that the renormalization group flow of the running couplings cross this
stability curve. Using the 1-loop approximation with fixed Yukawa coupling,
Chivukula et al.$^{[9]}$ found that, depending on the initial choice of
couplings, the trajectories either \\
\indent\indent  (i)~run to the infrared quasi fixed point near the origin,\\
\indent\indent (ii)~cross the stability line signaling a first order
transition \\
\indent\indent(iii)~or simply run away in which case the model is ill defined.

For example, focusing on the region of coupling space corresponding to
the initial couplings $\lambda_2(0)$ and $g^2(0)$
both large and positive and $\lambda_1(0)=0$,
the transition goes first order for sufficiently large
$\lambda_2(0)/g^2(0)$ ratio ($\stackrel{>}{\sim} 7$). Moreover, this occurs,
in general, near to the compositeness scale. For
example, choosing $\lambda_2(0)=10$ and $g^2(0)=1$ (and taking $N_C=3$), we
display in Fig. 1 the form of $Y(t)$ as a function of t. This function is seen
\begin{figure}
\vspace{2.5in}
\caption{$Y(t)$ as a function of $t$ for the initial parameters
$\lambda_1(0)=0, \lambda_2(0)=10, g^2(0)=1$
computed using the
1-loop renormalization group improved effective potential with fixed Yukawa
coupling.}
\end{figure}
to cross the stability curve $Y=0$ at $t_0\sim 1.3$. Thus the transition
turns first order and the hierarchy destabilizes
after $\sim 1.3$ e-foldings (which corresponds to $v\sim 0.27\Lambda$).
However, since the couplings are very large, the 1-loop perturbative
approximation can certainly be called into question. For instance, naively
using the 2-loop perturbative renormalization group
functions for these large initial couplings,
the renormalization group trajectories do not cross the line but simply run
away.

Clearly, some nonperturbative approximation scheme is required to
properly deal with the system in the vicinity of the strong coupling
compositeness scale. The purely bosonic $U(2)_L \times U(2)_R$
model (no chiral fermions) has been
simulated using lattice Monte Carlo techniques$^{[11]}$ and
was seen to undergo a Coleman-Weinberg instability.
An alternate approach which includes the chiral fermions
has been advocated by Bardeen et al.$^{[12]}$.
Modelling the nonperturbative physics in the vicinity of the compositeness
scale using a large $N_C$ approximation, then at scale $\Lambda$
the Yukawa coupling is seen to dominate. Retaining only it
and the fine tuned scalar mass term needed to cancel the additive quadratic
divergence, the model at scale $\Lambda$ reduces
to the minimal NJL model which is exactly soluble in the large $N_C$ limit
producing the running couplings
$\lambda_1(t)=0$ and
$\lambda_2(t)=3g^2(t)=\frac{48}{N_C t}$. This model exhibits
a (trivial) second order chiral
transition and as such allows any sized hierarchy
to be technically achieved. Note that for couplings consistent with
the large $N_C$ approximation, the 1-loop perturbative solution is driven to
the infrared quasi fixed point and thus also does
not cross the stability curve.
As such it too allows for arbitrary large hierarchies to be tuned.
Running the couplings using the large $N_C$ approximation solution
until the couplings have decreased sufficiently
to be smoothly joined onto
a perturbative running which also includes the running of the Yukawa coupling,
it is found that the stability line is eventually crossed but only after a
sizeable hierarchy ($t_0
\sim 20-25$) has been established. Note that for this case of initially
dominate Yukawa coupling (or for
one of comparable size to the initial scalar self
coupling), it is important to include the effects of its running which is
to enhance the flow of the system toward the stability line.
For smaller initial Yukawa couplings
the effects of its running will not be nearly
as important. Since the large $N_C$ approximation is nonperturbative,
the procedure of Bardeen et al. is a self consistent one. On the
other hand, it can be reliably employed for only a very limited range of the
initial parameter space.

An alternate nonperturbative method is provided by the continuous
Wilson renormalization group equation (WRGE)$^{[14-20]}$ which has been
extended to include chiral fermions$^{[17]}$. The WRGE
nonperturbatively relates the form
of the Euclidean action at a scale $e^{-t}\Lambda$ to the action at scale
$\Lambda$ for $t>0$. It is derived by demanding that the physics, ie.
correlation functions, remain unchanged as the degrees of freedom carrying
momentum between scales $\Lambda$ and $e^{-t}\Lambda$ are integrated out. Thus
either action can be used to equivalently describe the physics on all
scales less than $e^{-t}\Lambda$ and both actions lie on the same Wilson
renormalization group trajectory. The lower scale action is constructed by
appropriately changing the coefficients of the operators already present at
scale $\Lambda$, as well as including new ones,
in such a way so as to keep the physics unchanged. In general,
the resultant action incorporates a complete set of local operators. This
includes not only the relevant and marginal operators, but
also irrelevant ones. The coefficient of each operator is fixed in terms
of the initial action defined at scale $\Lambda$. The full WRGE is a very
complicated integro
functional differential equation and its analysis necessarily requires some
simplifying approximations. One commonly used approximation is to work
within a local action approximation$^{[16-17]}$
which ignores anomalous dimensions
and derivative interactions while we further neglect operators
higher than bilinear in the fermion fields. In addition, in order to obtain a
tractable analysis, we are forced to restrict attention to the case of a fixed
Yukawa coupling. As such, the initial parameter space that we are able to
investigate using this approach
is restricted to be one where the initial scalar self coupling(s) dominates
the initial Yukawa coupling which, however,
can still be of order unity. Thus the
non-perturbative WRGE we solve probes a region of initial parameter
space intermediate between the lattice
Monte Carlo simulations ($g(0)=0$) and the
large $N_C$ analysis ($g(0)$ dominates) and our results can be be viewed as
complementary to those analyses. Taking into
account the various simplifications,
the action at scale $e^{-t}\Lambda$ can be written as
$$  \everymath={\displaystyle}
   \begin{array}{rcl}
   S[\Sigma, \Sigma^{\dagger},\psi,\bar{\psi};t] &=& \int d^4 x
\{tr (\partial_\mu \Sigma^{\dagger} \partial^\mu \Sigma)
     + \bar{\psi} \gamma\cdot D\psi + V(x,y,t)\\
       && + \frac{\pi}{\sqrt{2}}g(0)(\bar{\psi}_L \Sigma \psi_R +
          \bar{\psi}_R \Sigma^{\dagger}\psi_L) \}  \, ,
     \end{array}
    \eqno{(6)}
$$
where $V(x,y,t)$ is the $U(2)_L\times U(2)_R$ invariant
potential function at this scale. The full WRGE$^{[17]}$ then
reduces to a partial differential
equation for this function of the form
$$ \everymath={\displaystyle}
   \begin{array}{rcl}
    \frac{\partial V}{\partial t} &=& 4V - \Sigma_{ij}
      \frac{\partial V}{\partial \Sigma_{ij}} -
      \Sigma_{ij}^{\dagger} \frac{\partial V}{\partial \Sigma_{ij}^{\dagger}}\\
       && + \frac{1}{16\pi^2} tr~\ell n (1 + W) - \frac{N_c}{4\pi^2}
       tr~\ell n (1 + \frac{\pi^2}{2}g^2(0) \Sigma^\dagger\Sigma),
    \end{array}
   \eqno{(7)}
$$
where $W(\Sigma, \Sigma^{\dagger})$ is the matrix of second derivatives
$$  \everymath={\displaystyle}
    W = \left[ \begin{array}{cc}
      \frac{\partial^2V}{\partial \Sigma_{ij}^{\dagger} \partial
\Sigma_{k\ell}} &
      \frac{\partial^2V}{\partial \Sigma_{ij}^{\dagger} \partial
\Sigma_{k\ell}^{\dagger}} \\
      \frac{\partial^2V}{\partial \Sigma_{ij} \partial \Sigma_{k\ell}} &
      \frac{\partial^2V}{\partial \Sigma_{ij} \partial\Sigma_{k\ell}^{\dagger}}
   \end{array} \right]
   \eqno{(8)}
$$
The $tr~\ell n$ terms containing $W$ arise from integrating out the scalar
modes,
while those containing $g^2(0)$ are due to integrating out the fermion modes.
After a considerable amount of algebraic manipulation, the various determinants
can be explicitly evaluated yielding the WRGE$^{[19]}$
$$ \everymath={\displaystyle}
   \begin{array}{rcl}
  &&\frac{\partial V}{\partial t} = 4V - 2xV_x - 4yV_y
 + \frac{1}{8\pi^2} \ell n[(1+V_x + 2xV_y)^2  - 2(x^2-y)V_y^2]\\
   &&\qquad + \frac{1}{16\pi^2} \ell n [(1+V_x)^2 + 2x(1+V_x)V_y
+ 2 (x^2-y)V_y^2]\\
&&+ \frac{1}{16\pi^2} \ell n [(1+V_x)(1+V_x
    + 6xV_y+ 2xV_{xx} + 8y V_{xy}+ 4x(3y-x^2)V_{yy})\\
   &&\qquad\qquad\qquad + 6(x^2-y)V_y (3V_y + 2V_{xx} + 4xV_{xy}
    + 4yV_{yy})\\
&&\qquad\qquad\qquad\qquad + 8(x^2-y)(x^2-2y)(V_{xy}^2 - V_{xx}V_{yy})]\\
&&\qquad - \frac{N_C}{4\pi^2}\ell n[1 + \frac{\pi^2}{2}xg^2 (0) +
\frac{\pi^4}{8} (x^2-y) g^4 (0)]\, ,
\end{array}
  \eqno{(9)}
$$
subject to the initial condition of Eq.(2). Here
the subscripts denote differentiation with respect to that variable so
that, for example, $V_x = \frac{\partial V}{\partial x}$ etc.
For $t\geq 0$, each action constructed using the
$V(x,y,t)$ satisfying this equation lies on the same Wilson
renormalization group trajectory and produces the same physics on all scales
less than $e^{-t}\Lambda$.
Unfortunately, the solution to this equation is currently beyond our
numerical abilities. Thus we make the further truncation
of retaining terms only up to linear in $y$ with coefficients which are
arbitrary functions of $x$. Eq.~(9) then reduces
to two coupled equations which are of a similiar (although
considerably
more complicated) form to what we previously solved in obtaining
nonperturbative mass bounds$^{[17,20]}$. While the
truncations used are drastic and uncontrolled, they still
include contributions from an infinite number of operators.

The resulting equations are then
numerically solved for $t$ values up to some $t^*$, where
$t^*$ lies in a region where $V(x,y,t)$ is found to be linear in
$t-t^*$ with a slope of the same form as the linearized in $t-t^*$
full 1-loop effective potential. The radiative
corrections arising from the momentum modes less than $e^{-t^*}\Lambda$
can then be satisfactorially incorporated using the 1-loop
approximation to the effective potential. Note that this 1-loop effective
potential also includes an infinite number of operators
as is necessary to allow a smooth joining to the WRGE solution. So doing,
the full effective potential is then constructed as
$$
V_{eff}(x,y)=V(x,y,t^*)+V_{1-loop}(x,y,t^*) \, ,
\eqno{(10)}
$$
where $V_{1-loop}(x,y,t^*)$
is the analytically calculable 1-loop effective potential
which accounts for the effects of the degrees of freedom carrying
momentum less than
$e^{-t^*}\Lambda$. A Coleman-Weinberg instability is signalled
by a non-trivial global minimum of $V_{eff}(v)=V_{eff}(x,y)|_{y=\frac{1}{2}
x^2=\frac{1}{2}v^4}$ with vanishing renormalized mass. The condition of
vanishing renormalized mass, $V^{eff}_x|_{x=y=0}$, is
achieved by an appropriate tuning of the parameter $m^2(0)$ appearing in
Eq.~(2). If such a minimum
appears at $v=e^{-t_0}\Lambda$, then the system can sustain a hierarchy only
over $t_0$ e-foldings.

Focusing on the specific initial couplings of $\lambda_1(0)=0,~
\lambda_2(0)=10,~ g^2(0)=1$, we numerically integrated the WRGE and found
that for $t^*\sim 1.5$,
$V(x,y,t)$ was linear in $t-t^*$ and smoothly joined
onto the 1-loop effective potential. As an indication of this, we plot in
Figs. 2-4 the behavior of the coefficients, $\lambda_1(t)=\frac{6}{\pi^2}
V^{eff}_{xx}|_{x=y=0},
\lambda_2(t)=\frac{6}{\pi^2}V^{eff}_{y}|_{x=y=0}$ and
$\ell(t)=e^{-2t}(\frac{6}{\pi^2})^2V^{eff}_{xy}|_{x=y=0}$,
of the two independent dimension four operators
and one of the dimension six operators as a function of $t$. Note that
we have included the explicit factor of $e^{-2t} \sim \frac{1}{\Lambda^2}$
accompanying the dimension six operator in the definition of its coupling.
\begin{figure}
\vspace{2.5in}
\caption{$\lambda_1(t)$ as a function
of $t$ computed using the nonperturbative WRGE with fixed Yukawa coupling.}
\end{figure}
\begin{figure}
\vspace{2.5in}
\caption{$\lambda_2(t)$ as a function
of $t$ computed using the nonperturbative WRGE with fixed Yukawa coupling.}
\end{figure}
As is clearly
demonstrated, these exhibit a linear behavior in $t-t^*$ for $t^*\sim 1.5$.
The behavior of these couplings in this region
is identical to that obtained using the linearized in $t-t^*$  1-loop
effective potential with anomalous dimensions neglected. Note that for
small $t$ values, the coefficient of the induced
dimension six operator (and other irrelevant operators) are quite
sizeable and play an important role in the dynamics. As $t$ increases,
the effects of the irrelevant operators is diminished and the system is
eventually attracted
to the space spanned by the relevant and marginal operators only.
Thus the contribution of the degrees of freedom
with momentum less than $e^{-t^*}\Lambda$ can then be included
using the 1-loop approximation.
\begin{figure}
\vspace{2.5in}
\caption{$\ell(t)$ as a function
of $t$ computed using the nonperturbative WRGE with fixed Yukawa coupling.}
\end{figure}

Since the
WRGE approach includes an infinite number of operators, it would require
an infinite
dimensional space to plot the renormalization group flows. Thus we focus
instead directly on the vacuum effective potential,
$V_{eff}(v)$, defined in Eq.~(10). In Fig. 5, we plot the generalized
Yamagishi function $Y(t)=-\frac{12}{\pi^2}e^{4t}\frac{dV_{eff}}{dt}$
as a function of
$t=-\ell n~ v$.  A first order transition is signaled if this function crosses
the stability line $Y=0$.
\begin{figure}
\vspace{2.5in}
\caption{$Y(t)$ as a function of $t$ for the initial parameters
$\lambda_1(0)=0, \lambda_2(0)=10, g^2(0)=1$
computed using the nonperturbative WRGE with fixed Yukawa
coupling.}
\end{figure}
It is seen that such a zero occurs at $t_0 \sim 2.0$ and corresponds to
a non-trivial global minimum of the effective potential at $v\sim 0.14\Lambda$.
The value of $t_0$ was seen to be insensitive to changes in the choice of
$t^*$. Thus the transition goes
first order and a hierarchy of only $\sim 2.0$
e-foldings can be established.
This result is in qualitative agreement with that found using
the 1-loop perturbative approximation.

We have also investigated the phase structure of the model for other values
of the initial coupling parameter space using the WRGE with
fixed $g^2(0)$. For instance, for
$\lambda_1(0)=-2,\lambda_2(0)=10, g^2(0)=1$,
the Yamagishi function $Y(t)$ crosses the
stability line at $t_0=1.9 $ indicating a first order transition
after 1.9 e-foldings.
Once again, this is in qualitative agreement with the 1-loop
perturbative result of $t_0=1.0$. Similarly, for
$\lambda_1(0)=2,\lambda_2(0)=10, g^2(0)=1$, we find the chiral symmetry
breaking transition turns first order at $t_0=2.1$ while the
1-loop perturbative approximation gives $t_0=1.7$.

Using the nonperturbative WRGE, we have studied the stability of fine tuned
hierarchies in a $U(2)_L\times U(2)_R$ chiral model for a range of initial
couplings satisfying $\lambda_2(0)>>g^2(0)\sim |\lambda_1(0)|\sim 1$. In
general, we found that
the chiral symmetry phase transition turns first order in close proximity to
the compositeness scale. This is in qualitative agreement with the results
obtained using the 1-loop improved effective potential. In performing our
analysis, we were forced to employ various truncations in order to render the
numerical integrations tractable. Clearly the sensitivity of our results to
these various approximations needs further scrutiny. Moreover, some of these
truncations must be eliminated before we are able to probe the
region of initial parameter space with larger Yukawa couplings. We hope to
address these issues in future studies.
\\

\bigskip

We thank Sergei Klebnikov for many enjoyable useful discussions and Bijan
Haeri for assistance with the numerical computations. This work
was supported in part by the U.S. Department of Energy under
grant DE-AC02-76ER01428 (Task B).

\vspace{0.2in}
\noindent
{\bf References}

\singlespace
\begin{enumerate}

\item
T. Appelquist, T. Takeuchi, M. Einhorn and L.C.R. Wijewardhana, Phys. Lett.
{\bf 220B} (1989) 223;
T. Takeuchi, Phys. Rev. {\bf D20} (1989) 2697;
V.A. Miransky and K. Yamawaki, Mod. Phys. Lett. {\bf A4} (1989) 129.

\item
C.T. Hill, D.C. Kennedy, T. Onogi and H.-L. Yu, Phys. Rev. {\bf D 47} (1993)
2940.

\item
Y. Nambu, in the {\bf Proceedings of the XI International Symposium on
Elementary Particle Physics}, Kazimierz, Poland, 1988, ed. by Z. Adjuk et al.,
(World Scientific, 1989); Enrico Fermi Insitute preprint EFI 89-08
(unpublished).

\item
V.A. Miransky, M. Tanabashi and K. Yamawaki, Phys. Lett.
{\bf 221B} (1989) 177; Mod. Phys. Lett. {\bf A7} (1989) 1042.

\item
W.A. Bardeen, C.T. Hill and M. Lindner, Phys. Rev. {\bf D 41} (1990) 1647.

\item
C.T. Hill M. Luty and E.A. Paschos, Phys. Rev. {\bf D43} (1991) 3011;
C.T. Hill, Phys. Lett. {\bf 266B} (19991) 419;
S. Martin, Phys. Rev. {\bf D45} (1992) 4283 and
Phys. Rev. {\bf D46} (1992) 2197;
T. Eliot and S.F. King, Phys. Lett. {\bf 283B} (1992) 371;
M. Lindner and D. Lust, Phys. Lett. {\bf 272B} (1991) 91;
M. Lindner and D. Ross, Nucl. Phys. {\bf B370} (1992) 30;
N. Evans, S. King and D. Ross, Z. Phys. {\bf C 60} (1993) 509.

\item
Y. Nambu and G. Jona-Lasinio, Phys. Rev. {\bf 122} (1961) 345.

\item
S. Coleman and E. Weinberg, Phys. Rev. {\bf D7} (1973) 1888.

\item
R.S. Chivukula, M. Golden and E.H. Simmons, Phys. Rev. Lett. {\bf 70} (1993)
1587.

\item
F. Abe et al. (CDF Collaboration) Phys. Rev. Lett. {\bf 73} (1994) 225;
Phys. Rev. {\bf D50} (1994) 2966.

\item
Y. Shen, Phys. Lett. {\bf 315B} (1993) 146.

\item
W.A. Bardeen, C.T. Hill and D. Jungnickel, Phys Rev. {\bf D49} (1994) 1437.

\item
H. Yamagichi, Phys. Rev. {\bf D23} (1981) 1880.

\item
K.G. Wilson, Phys. Rev. {\bf B4} (1971) 3174, 3184;
K.G. Wilson and J.B. Kogut, Phys. Rept. {\bf 12C} (1974) 75;
K.G. Wilson, Rev. Mod Phys. {\bf 55} (1983) 583.

\item
F.J. Wegner and A. Houghton, Phys. Rev. {\bf A8} (1973) 401;
J. Polchinski, Nucl. Phys. {\bf B231} (1984) 269;
S. Weinberg, in {\bf Proceedings of the 1976 International School of
Subnuclear Physics}, Erice, ed. by A. Zichichi (Plenum Press, 1978).

\item
A. Hasenfratz and P. Hasenfratz, Nucl. Phys. {\bf B270} (1986) 685;
P. Hasenfratz and J. Nager, Z. Phys. {\bf C37} (1988) 477.

\item
T.E. Clark, B. Haeri and S.T. Love, Nucl. Phys {\bf B402} (1993) 628.

\item
U. Ellwanger and L. Vergara, Nucl. Phys. {\bf B398} (1993) 52;
C. Wetterich, Int. J. Mod. Phys. {\bf A9} (1994) 3571;
T. Morris, Int. J. Mod. Phys. {\bf A9} (1994) 2411 and Phys. Lett. {\bf 334B}
(1994) 355;
M. Alford and J. March-Russell, Nucl. Phys. {\bf B417} (1994) 527.

\item
S.T. Love, PURD-TH-94-09, to appear in the {\bf Proceedings of the XXVII
International Conference on High Energy Physics}, Glasgow 1994.

\item
T.E. Clark, B. Haeri, S.T. Love, M.A. Walker and W.T.A. ter Veldhuis,
Phys. Rev. {\bf D50} (1994) 606; S.T. Love in {\bf Proceedings of the
International Europhysics Conference on High Energy Physics}, Marseilles 1993
ed. J. Carr and M. Perrottet (Editions Frontiere, 1994) 208.

\end{enumerate}

\end{document}